\documentclass[aps,prd,preprintnumbers,nofootinbib,floatfix,11pt]{revtex4}
\usepackage{amssymb}
\usepackage{amsmath,amscd}
\usepackage{color}
\usepackage{mathrsfs}

\usepackage{bm}
\usepackage[normalem]{ulem}
\usepackage{accents}

\newcommand{\bcen}{\begin{center}}
\newcommand{\ecen}{\end{center}}
\newcommand{\bflr}{\begin{flushright}}
\newcommand{\eflr}{\end{flushright}}
\newcommand{\bfll}{\begin{flushleft}}
\newcommand{\efll}{\end{flushleft}}
\newcommand{\beq}{\begin{equation}}
\newcommand{\eeq}{\end{equation}}
\newcommand{\beqa}{\begin{eqnarray}}
\newcommand{\eeqa}{\end{eqnarray}}
\newcommand{\bite}{\begin{itemize}}
\newcommand{\eite}{\end{itemize}}
\newcommand{\benu}{\begin{enumerate}}
\newcommand{\eenu}{\end{enumerate}}

\newcommand{\ep}[0]{ {\epsilon} }

\newlength{\dhatheight}
\newcommand{\doublehat}[1]{%
    \settoheight{\dhatheight}{\ensuremath{\hat{#1}}}%
    \addtolength{\dhatheight}{-0.35ex}%
    \hat{\vphantom{\rule{1pt}{\dhatheight}}%
    \smash{\hat{#1}}}}

\newcommand{\cO}{{\cal O}}
\newcommand{\cOb}[1]{{{\cal O} \left( {#1} \right)}}
\newcommand{\eq}[1]{(\ref{#1})}

\newcommand{\D}{{\cal D}}

\newtheorem{theorem}{Theorem}
\newtheorem{definition}{Definition}

\begin{document}

\title{Attractive gravity probe surfaces in higher dimensions}

\author{Keisuke Izumi${}^{1,2}$, Yoshimune Tomikawa${}^3$, Tetsuya Shiromizu${}^{2,1}$, Hirotaka Yoshino${}^4$}

\affiliation{${}^1$Kobayashi-Maskawa Institute, Nagoya University, Nagoya 464-8602, Japan}
\affiliation{${}^2$Department of Mathematics, Nagoya University, Nagoya 464-8602, Japan}
\affiliation{${}^3$Division of Science, School of Science and Engineering, Tokyo Denki University, Saitama 350-0394, Japan}
\affiliation{${}^4$Department of Physics, Osaka Metropolitan University, Osaka 558-8585, Japan}

\begin{abstract}
A generalization of the Riemannian Penrose inequality in $n$-dimensional space ($3\le n<8$) is done. 
We introduce a parameter $\alpha$ ($-\frac{1}{n-1}<\alpha < \infty$) indicating the strength of the gravitational field, 
and  define a refined attractive gravity probe surface (refined AGPS) with $\alpha$. 
Then, we show the area inequality for a refined AGPS, 
$A \le \omega_{n-1} \left[ (n+2(n-1)\alpha)Gm /(1+(n-1)\alpha) \right]^{\frac{n-1}{n-2}}$, 
where $A$ is the area of the refined AGPS,  $\omega_{n-1}$ is the area of the standard unit $(n-1)$-sphere, $G$ is 
Newton's gravitational constant and $m$ is the Arnowitt-Deser-Misner mass. 
The obtained inequality is applicable not only to surfaces in strong gravity regions such as a minimal surface 
(corresponding to the limit $\alpha \to \infty$), 
but also to those in weak gravity existing near infinity (corresponding to the limit $\alpha \to -\frac{1}{n-1}$) . 
\end{abstract}

\maketitle

\section{Introduction}

Study of surfaces is one of the ways to investigate the gravitational theory. 
For instance, an event horizon, defined as a boundary of the causal past of future null infinity, 
shows the distinctive properties of the gravitational theory, such as its area increasing law~\cite{Hawking:1971tu}. 
Especially, the black hole thermodynamics is expected as one of the keys to the quantum gravity, 
and the area of the event horizon is interpreted as the entropy of the black hole~\cite{Bekenstein:1972tm,Bekenstein:1973ur}.
The notion of the entropy is extended to the Ryu-Takayanagi surface~\cite{Ryu:2006bv,Ryu:2006ef} in the AdS/CFT correspondence, 
and it is applied to the quantum information theory~\cite{Nishioka:2018khk}. 

The Penrose inequality~\cite{PI}, which is the main topic of this paper, is a conjecture about apparent horizons. 
The conjecture states that the area of a trapped surface has the upper bound characterized by that of the 
Schwarzschild solution with the same Arnowitt-Deser-Misner (ADM) mass~\cite{Arnowitt:1959ah,Arnowitt:1960,Arnowitt:1962hi}, 
that is $A_{\rm AH}\le 4\pi (2Gm)^2$ in four-dimensional spacetime, where $A_{\rm AH}$ is the area of the apparent horizon, $G$ 
is Newton's gravitational constant and $m$ is the ADM mass. 
It is expected to hold if the cosmic censorship conjecture is true. 
The Penrose inequality was proved in special situations: 
on a time-symmetric hypersurface (under the assumption of its existence) the inequality was shown~\cite{J&W,H&I,Bray}. 
This statement on a time-symmetric hypersurface is equivalent to that 
the area of an outermost minimal surface  $A_{\rm MS}$ in asymptotically flat space with nonnegative Ricci scalar is bounded as $A_{\rm MS}\le 4\pi (2Gm)^2$, which is called the Riemannian Penrose inequality.
The Geroch energy~\cite{Geroch}, 
which had been originally introduced to show the positivity of the energy including gravity, was used in the proof~\cite{J&W,H&I}, 
and thus, the relation to the positive energy theorem~\cite{Schon:1979rg,Witten:1981mf} is one of the interesting subjects of study. 
Furthermore, the analysis with the Geroch energy was generalized to spaces with negative cosmological constants~\cite{Boucher:1983cv,Gibbons:1998zr}, and applied to the study of the Ryu-Takayanagi surface in the AdS/CFT 
correspondence~\cite{Fischetti:2016fbh}. 

Recently, generalizations of the Penrose inequality applicable to regions with weaker gravitational field have been achieved in three dimensional spaces~\cite{Shiromizu:2017ego,Izumi:2021hlx}. 
Its motivations are as follows. 
Since apparent horizons exist in black holes under the  cosmic censorship conjecture, 
they are not observable objects for the distant observers. 
Thus, the Penrose inequality is never be verified in observations. 
In order to improve this circumstance, the authors of this paper have introduced the generalization of photon sphere, 
which is one of the important objects in black hole observations, named the loosely trapped surface (LTS)~\cite{Shiromizu:2017ego}, and have shown its area inequality. 
The generalizations of the Penrose inequality are also important in the theoretical point of view. 
The outside regions of event horizons provide us sometimes main stage of  study in gravitational theory (especially, in quantum gravity 
including the Hawking radiation), but the trapped surface is located inside black holes. 
The generalizations to weak gravity region enable us to apply the inequality to such studies. 
The authors of this paper have given the further generalization of the Riemannian Penrose inequality~\cite{Izumi:2021hlx}. 
They have introduced attractive gravity probe surfaces (AGPSs), which can exist near the infinity, 
and shown its area inequality. 
This result would permit us to understand the meaning of the area inequality in the Newtonian approximation, and 
to relate the inequality with the asymptotic structures of spacetime, such as the Bondi-Metzner-Sachs symmetry~\cite{Bondi:1962px,Sachs:1962wk}. 

The higher-dimensional version of the Riemannian Penrose inequality was proved by Bray and Lee~\cite{Bray:2007opu}. 
In this paper, we show the area inequality of attractive gravity probe surfaces in spaces whose dimension is higher 
than or equal to three (but less than eight%
\footnote{
Since our proof relies on Bray and Lee's theorem~\cite{Bray:2007opu} using  Schoen and Yau's positive energy theorem~\cite{Schon:1979rg}, 
which is true only for $n$-dimensional space with $n<8$, our theorem holds under the same condition. 
However, Schoen and Yau gave a report on the establishment of the positive energy theorem in any dimensions~\cite{Schon:2017}, 
which may enable our theorem to apply in any dimensions. 
}), following the proof done for three dimensional space in our previous paper~\cite{Izumi:2021hlx}. 
We also have other progress in addition to the extension to higher-dimensional cases. 
The condition in the definition of the AGPS is refined, which includes all original AGPSs. 
Hence, the inequality is applied to more surfaces. 
Another progress is relaxing the conditions of the theorem. 
In the previous paper~\cite{Izumi:2021hlx}, we assumed the existence of the local inverse mean curvature flow, 
which requires information on the region near the AGPS ({\it i.e.} other than the AGPS), 
while in the proof of this paper we succeed to remove this condition. 

This paper is organized as follows.
In Sec. \ref{defAGPS}, we give the definition of refined AGPSs. 
The main theorem is shown in Sec. \ref{MT}.
The proof is presented in Secs. \ref{SoP}-\ref{comp}. 
In Sec.~\ref{SoP}, we give the idea of the proof. 
Then, we find that two conditions are required to be shown: 
the existence of a sequence of smooth manifolds approaching  the manifold constructed in Sec.~\ref{SoP} 
and nonexistence of the minimal surface anywhere but the boundary of the constructed manifold. 
The former is fixed in Sec.~\ref{SE}, while the latter is in Sec.~\ref{OM}. 
The proof is completed in Sec.~\ref{comp}.
Section~\ref{Summary} is devoted to a summary and discussion.  
Appendix~\ref{App} is given to show the existence of functions introduced for the discussion in Sec.~\ref{SE}.

We here show our notations of {\it inward} and {\it outward}, which would be different from the usual definitions. 
A manifold $\Sigma$ with boundaries, one of which is the infinity and the others of which are (refined) AGPSs, is our interest of study. 
We define {\it inward} and {\it outward} of an AGPS as follows. 
If a vector on an AGPS directs into $\Sigma$, we call the direction of the vector ``{\it outward}'' 
because it matches with the physical intuition that the vector naively directs to the infinity. 
The opposite direction is called ``{\it inward}''. 
The words ``{\it outside}'' is used in the similar way, that is, the region exists in the {\it outward} direction is called {\it outside}. 

\section{Refined Attractive Gravity Probe Surfaces}
\label{defAGPS}

In our previous paper~\cite{Izumi:2021hlx}, for proving an area inequality in three-dimensional spaces, 
the notion of attractive gravity probe surfaces (AGPSs) is introduced, 
which are defined as surfaces satisfying the positivity of the mean curvature $k>0$ and 
\begin{eqnarray}
r^a D_a k \ge \alpha k^2, \label{previous}
\end{eqnarray}
where $r^a$ is the outward unit normal, $D_a$ is the covariant derivative of the three-dimensional space and $k$ is defined 
by $k=D^ar_a$.  
In this paper, the area inequality will be generalized into higher-dimensional cases and the definition of AGPS will be refined. 

Let us investigate the geometric identity (for example, see Refs. \cite{Shiromizu:2017ego, Izumi:2021hlx}), 
\begin{eqnarray}
r^a D_a k + \frac12 {}^{(n)}R + \frac12  \mathsf{ k}_{ab}^2 
= \frac12 {}^{(n-1)}R - \varphi^{-1} \D^2 \varphi - \frac{n}{2(n-1)} k^2,
\label{ident}
\end{eqnarray}
where $\mathsf{ k}_{ab}$ and $\varphi$ are the traceless part of the extrinsic curvature $k_{ab}$ and the 
lapse function, respectively, and 
$\mathsf{ k}_{ab}^2$ means $\mathsf{ k}_{ab}\mathsf{ k}^{ab}$. ${}^{(n)}R$ is the Ricci scalar of $n$-dimensional space, and 
${}^{(n-1)}R$ and ${\cal D}_a$ are the Ricci scalar and the covariant derivative of $(n-1)$-dimensional hypersurface. 
Since manifolds with the nonnegative Ricci scalar ${}^{(n)}R \ge 0$ are our interest, 
the second term in the left-hand side is nonnegative. 
The third term 
is manifestly nonnegative. 
Thus, the condition of  Eq.~(\ref{previous}) with replacing $r^a D_a k$ by the left-hand side of Eq.~(\ref{ident}) 
becomes weaker than that original one. 
If the local inverse mean curvature flow ($\varphi k =1$) is assumed, 
 the right-hand side of Eq.~(\ref{ident}) is written as
\begin{eqnarray}
\frac12 {}^{(n-1)}R + k^{-1} \D^2 k -2 k^{-2} \left( \D k\right)^2 - \frac{n}{2(n-1)} k^2.\label{ident-rhs}
\end{eqnarray}
This is expressed in terms of the geometrical quantities of the $(n-1)$-dimensional hypersurface 
and the mean curvature, that is, these quantities is uniquely fixed by how the $(n-1)$-dimensional hypersurface is embedded.
Meanwhile, $r^a D_a k$ includes the second derivative of the induced metric, which 
requires more information than the embedding.
Thus, through Eqs. (\ref{ident}) and (\ref{ident-rhs}), there is a room to improve the condition \eqref{previous}. 
This motivates us to refine the conditions of AGPS as follows.
\noindent
%
\begin{definition}{\bf Refined Attractive Gravity Probe Surfaces (Refined AGPSs) :\ }
  Suppose $\Sigma$ to be a smooth $n$-dimensional manifold with
  a positive definite metric $g$.
A smooth compact hypersurface $S_\alpha$ in $\Sigma$ is a refined attractive gravity probe surface (refined AGPS) with a parameter $\alpha$ ($\alpha>-\frac{1}{n-1}$)
if the following conditions are satisfied everywhere on $S_\alpha$: 
\begin{description}
\setlength{\leftskip}{5mm}
 \item[(i)]\ The mean curvature $k$ is positive.
 \item[(ii)]\ The inequality 
\begin{eqnarray}
{}^{(n-1)}R + 2k^{-1} \D^2 k -4 k^{-2} \left( \D k\right)^2\ge \left( 2 \alpha + \frac{n}{n-1}\right) k^2  \label{rDkineq}
\end{eqnarray}
is satisfied. %
\footnote{
It may be better to introduce a new parameter equal to $\left( 2 \alpha + \frac{n}{n-1}\right)$ for simplification. 
However, we use $\alpha$ because it makes the relation to our previous study in Ref.~\cite{Izumi:2021hlx} clear.
}
\end{description}
Here, ${}^{(n-1)}R$, $k$ and $\D$ are the Ricci scalar, the mean curvature and covariant derivative of $S_\alpha$.
Surface $S_\alpha$ is not required to be connected. It can be multiple.
\label{definition-AGPS}
\end{definition}
%
%
\vspace{2mm}
Note that the conditions are written only with the functions and operators of the $(n-1)$-dimensional hypersurface. 
For original AGPSs defined in our previous paper~\cite{Izumi:2021hlx} the existence of the local inverse mean curvature flow 
near the $(n-1)$-dimensional hypersurface is required to be imposed, 
whereas the current definition of refined AGPSs is written only with the variables fixed by the embedding and does not require any other details of geometrical information around AGPSs.  
We will prove the area inequality with this refined condition in $n$-dimensional ($n\ge 3$) asymptotically flat spaces.

\section{Main Theorem} \label{MT}
Now, we present our main theorem.
%
\begin{theorem}
Let $\Sigma$ be an asymptotically flat, $n$-dimensional ($3\le n <8$), smooth  manifold with nonnegative Ricci scalar. 
The boundaries of $\Sigma$ are composed of an asymptotically flat end and
a refined AGPS $S_\alpha$, which can have multiple components,
with a parameter $\alpha$. 
Here, the unit normal $r^a$ to define the mean curvature is taken to be outward of $S_\alpha$.
Suppose that there exists a finite-distance smooth extension of $\Sigma$ from $S_\alpha$ to the interior of $S_\alpha$
satisfying the nonnegativity of the Ricci scalar.
Moreover, suppose that no minimal hypersurface satisfying either one of the following conditions exists:
\begin{description}
\setlength{\leftskip}{5mm}
 \item[(i)] \ It encloses (at least) one component of $S_\alpha$. 
 \item[(ii)] \  It has boundaries on $S_\alpha$ and
   its area is less than  $\omega_{n-1} (2Gm)^{\frac{n-1}{n-2}}$, where $\omega_{n-1}$ is the area of the standard unit $(n-1)$-sphere. 
\end{description}
Then, the area of $S_\alpha$ has an upper bound, 
\begin{eqnarray}
A_{\alpha} \le \omega_{n-1} \left[ \frac{n+2(n-1)\alpha}{1+(n-1)\alpha} Gm\right]^{\frac{n-1}{n-2}},
\label{ineqtheorem}
\end{eqnarray}
where $m$ is the ADM mass of the manifold and
$G$ is Newton's gravitational constant.
Equality holds if and only if $\Sigma$ is a time-symmetric hypersurface
of a Schwarzschild spacetime and 
$S_\alpha$ is a spherically symmetric hypersurface with  $r^a D_a k = \alpha k^2$. 
\label{theorem}
\end{theorem}
%
\vspace{2mm}
\noindent
Note that any AGPS, defined in the previous paper~\cite{Izumi:2021hlx}, satisfies the condition of refined AGPS. 
Thus, Theorem~\ref{theorem} holds for any AGPS.  
The proof of the theorem is given in the following sections.
Even if no finite-distance smooth extension of $\Sigma$ with the nonnegativie Ricci scalar exists, 
we can show the inequality~\eqref{ineqtheorem} if,  in a neighborhood of $S_\alpha$ on $\Sigma$, there exists a 
smooth foliation of hypersurfaces all of which (but $S_\alpha$) satisfy conditions 
$(i)$ and $(ii)$. 
We do not show the proof because it can be done by following the procedure shown in our 
previous paper~\cite{Izumi:2021hlx}. 
We emphasize again that the proof based on this paper does not require the local mean curvature flow in the 
neighborhood of $S_\alpha$, which is one of the major revisions from our previous paper~\cite{Izumi:2021hlx}.

\section{Sketch of Proof}\label{SoP}

We take a smooth normal coordinate near $S_\alpha$ as
\begin{eqnarray}
ds^2= \varphi^2 dr^2 + g_{ab}dx^a dx^b \label{nonm}
\end{eqnarray} 
such that $S_\alpha$ is located at $r=0$, $\varphi k=1$ is imposed only  on $S_\alpha$ and the manifold $\Sigma$ exists in $r\ge 0$. 
Note that condition $\varphi k =1 $ is imposed only on $S_\alpha$, and thus the existence of the local inverse mean curvature flow is not assumed.
The positivity of $k$ on $S_\alpha$ implies that $\varphi$ is positive.

Let us introduce a manifold $\bar \Sigma$
with a metric in the range $r\le0$ 
\begin{eqnarray}
&&d\bar s^2 
=  \frac{ 1-\exp \left( -\frac{n-2}{n-1}r_0 \right)}{1-\exp \left[ -\frac{n-2}{n-1}(r+r_0) \right]}\exp \left( \frac{2}{n-1} r \right) \varphi_0^2 (x^a) dr^2 
\nonumber \\
&& \hspace{50mm}
+\exp \left( \frac{2}{n-1}r \right) g_{0,ab} (x^a) dx^a dx^b,
\label{barm}
\end{eqnarray}
where  $g_{0,ab}$ is the metric of $S_\alpha$, $\varphi_0$ is taken as 
\begin{eqnarray}
\varphi_0(x^a):=\varphi|_{r=0}  \left(=k^{-1}|_{S_\alpha} \right) \label{vp0}
\end{eqnarray}
and $r_0$ is defined as a constant  satisfying
\begin{eqnarray}
\exp \left(\frac{n-2}{n-1}r_0\right) = \frac{n+ 2(n-1)\alpha}{2[1+ (n-1)\alpha]}.
\label{r0}
\end{eqnarray}
The manifold $\bar\Sigma$ is continuously glued to $\Sigma$ at $r=0$, because all components of both induced metrics \eqref{nonm} and \eqref{barm} on the gluing  
surface $S_\alpha$ $(r=0)$ are the same. 
However, their extrinsic curvatures do not necessarily match with each other and then 
the metric components are generally $C^0$-class there. 

Since, for $\bar \Sigma$, the metric components for each $r$-constant hypersurface are written by the product of 
$r$-dependent part and $x^a$-dependent part,  each hypersurface becomes umbilical, that is, the extrinsic curvature is written as
\begin{eqnarray}
&& \bar k_{ab} = \frac1{n-1} \bar k \bar g_{ab}. 
\end{eqnarray}
Hereinafter, quantities with bars indicate that they are associated
with the metric \eqref{barm}.
The mean curvature and its normal derivative of the metric~\eq{barm} are calculated as
\begin{eqnarray}
&& \bar k = \varphi_0^{-1} \sqrt{\frac{1-\exp \left[ -\frac{n-2}{n-1}(r+r_0) \right]}{ 1-\exp \left( -\frac{n-2}{n-1}r_0 \right)}\exp \left(- \frac{2}{n-1}r \right)}, \\
&&\bar{r}^a \bar{D}_a \bar{k} = -\frac1{n-1} \bar k^2 \frac{1-\frac{n}{2} \exp \left[ -\frac{n-2}{n-1} (r+r_0) \right] }{1- \exp \left[ - \frac{n-2}{n-1}(r+r_0) \right] },
\end{eqnarray}
where $\bar{r}^a = \bar \varphi^{-1} ( \partial/ \partial  r)^a$.
Note that  $\bar k|_{r=0} = \varphi_0^{-1}= k|_{S_\alpha}$ and $\bar{r}^a \bar{D}_a \bar{k}|_{r=0} =  \alpha \bar k^2|_{r=0}
$ are satisfied, where we use Eq.~(\ref{r0}).

The $n$-dimensional Ricci scalar of $\bar\Sigma$ is expressed with the $(n-1)$-dimensional quantities  as
\begin{eqnarray}
{}^{(n)}\bar R =  {}^{(n-1)}\bar R - 2 \bar \varphi^{-1} \bar {\cal D}^2 \bar \varphi - 2 \bar{r}^a \bar{D}_a \bar{k} - \frac{n}{n-1} \bar k^2,
\label{SPbarR}
\end{eqnarray} 
where we define $\bar \varphi$ as the lapse function in the metric~\eq{barm}, that is,
\begin{eqnarray}
\bar \varphi :=  \varphi_0 (x^a) \sqrt{ \frac{ 1-\exp \left( -\frac{n-2}{n-1}r_0 \right)}{1-\exp \left[ -\frac{n-2}{n-1}(r+r_0) \right]}\exp \left(\frac{2}{n-1} r \right) }.
\end{eqnarray} 
With the explicit form of the metric~\eq{barm},
all terms in the right-hand side of Eq.~(\ref{SPbarR}) are written with those on $S_{\alpha}$,
\begin{eqnarray}
&& {}^{(n-1)}\bar R = \exp \left(-\frac{2}{n-1}r\right) {}^{(n-1)} R_0,  \qquad   
 \bar \varphi^{-1} \bar {\cal D}^2 \bar \varphi = \exp \left(-\frac{2}{n-1}r\right)  \varphi_0^{-1}  {\cal D}^2  \varphi_0, 
\nonumber \\
&&
   2 \bar{r}^a \bar{D}_a \bar{k} + \frac{n}{n-1} \bar k^2 = \exp \left(-\frac{2}{n-1}r\right) \left(2\alpha +\frac{n}{n-1} \right)\varphi_0^{-2},
\end{eqnarray} 
where each index "0" indicates the quantity on the $r=0$ hypersurface $S_\alpha$. They give 
\begin{eqnarray}
{}^{(n)}\bar R = \exp \left(-\frac{2}{n-1}r\right)\left\{ {}^{(n-1)} R_0 - 2 \varphi_0^{-1}  {\cal D}^2  \varphi_0 -  \left( 2\alpha +\frac{n}{n-1}  \right)\varphi_0^{-2}\right\}. 
\label{barnR}
\end{eqnarray} 
If $S_\alpha$ is a refined AGPS, Eq.~(\ref{barnR})  with Eqs.~(\ref{rDkineq}) and (\ref{vp0}) results in the nonnegativity of ${}^{(n)}\bar R$.

We have shown the nonnegativity of the $n$-dimensional Ricci scalar of $\bar\Sigma$, 
while that of $\Sigma$ is one of the assumptions of the theorem.
Therefore,  everywhere on the glued manifold $\Sigma \cup \bar\Sigma$ (but on $S_\alpha$) the $n$-dimensional Ricci scalar is nonnegative.  
Hence, one expects the application of Bray \& Lee's proof to the Riemannian Penrose inequality~\cite{Bray:2007opu}
for $\Sigma \cup \bar \Sigma$. 
However, their theorem requires the smoothness of the manifold as an assumption, 
whereas our manifold $\Sigma \cup \bar\Sigma$ is generally $C^0$-class, not $C^\infty$-class on $S_\alpha$.
This problem will be fixed in the following sections.
In the rest of the present section,
supposing that  Bray \& Lee's theorem is applicable to the manifold $\Sigma \cup \bar\Sigma$,
we show the inequality \eqref{ineqtheorem}. 

The extended manifold $\Sigma \cup \bar\Sigma$ has a minimal hypersurface at $r=-r_0$. 
Suppose that the $r=-r_0$ hypersurface is the outermost minimal hypersurface, which will be justified in Sec.~\ref{OM}. 
Since the metric of $\bar\Sigma$ is explicitly written in Eq.~\eqref{barm}, 
we can relate the areas of $S_{\alpha}$ and of the minimal hypersurface ${\mathscr S}_0$ at $r=-r_0$,
\begin{eqnarray}
A_0 = \exp (-r_0 ) A_\alpha =\left[\frac{2\{1+(n-1)\alpha\}}{n+2(n-1)\alpha}\right]^{\frac{n-1}{n-2}} A_\alpha,
\label{AalphaA0}
\end{eqnarray} 
where $A_0$ and $A_\alpha$ are the areas of ${\mathscr S}_0$ and $S_\alpha$, respectively. 
In the second equality, we used Eq.~\eqref{r0}. 
If Bray \& Lee's theorem is applicable for ${\mathscr S}_0$, the area of ${\mathscr S}_0$ is bounded as
\begin{eqnarray}
A_0 \le \omega_{n-1} (2Gm)^{\frac{n-1}{n-2}}.
\label{PE}
\end{eqnarray} 
This gives
\begin{eqnarray}
A_{\alpha} \le \omega_{n-1} \left[ \frac{n+2(n-1)\alpha}{1+(n-1)\alpha} Gm\right]^{\frac{n-1}{n-2}},
\label{MPE}
\end{eqnarray} 
that is, the inequality~\eqref{ineqtheorem} is obtained.

When the equality holds in inequality~\eq{MPE}, 
it also holds in the Riemannian Penrose inequality~\eq{PE}. 
Bray \& Lee's theorem implies that the manifold has the metric of the time-symmetric slice of
a (higher-dimensional) Schwarzschild spacetime. 
Then, the minimal hypersurface ${\mathscr S}_0$ is spherically symmetric, and thus $S_\alpha$ too because of the metric form 
of Eq.~\eqref{barm}. 
Moreover, by construction of $\bar \Sigma$, $r^a D_a k = \alpha k^2$ holds on $S_\alpha$.
As a result, the equality in the inequality \eq{MPE}
holds if and only if 
$\Sigma$ is the time-symmetric hypersurface of a Schwarzschild spacetime, and then, 
$S_\alpha$ is the spherically symmetric hypersurface with  $r^a D_a k = \alpha k^2$.

In the above discussion, it is reminded that two assumptions are imposed: 
${\mathscr S}_0$ is the outermost minimal surface, and Bray \& Lee's theorem is applicable for $\Sigma \cup \bar\Sigma$. 
The former is justified from the assumptions of the theorem, 
which we will see in Sec.~\ref{OM}.
For the latter, in Sec.~\ref{SE}, we will 
show 
that $\Sigma \cup \bar\Sigma$ is achieved as a limit of a sequence of  smooth manifolds for which 
Bray \& Lee's theorem is applicable. 

\section{Smooth extension}\label{SE}

In this section, we construct a sequence of  smooth manifolds and show that $\Sigma \cup \bar\Sigma$ is obtained as a limit of it. 
We carry it out in the following steps. 
From the assumption of the theorem, there exists a smooth extention of $\Sigma$ from $S_\alpha$ with nonnegative 
 Ricci scalar. 
At first in Sec. \ref{first}, we deform this extended region of $\Sigma$ so that the Ricci scalar is strictly positive.
We call the deformed manifold $\hat \Sigma$. 
Next, in Sec. \ref{second}, based on the manifold $\hat \Sigma$, we construct the $C^0$-class extension discussed in the previous section. 
This $C^0$-class extension is done on a surface slightly inward from $S_\alpha$, parameterized with a small positive value $\epsilon$. 
Due to the strict positivity of the Ricci scalar of $\hat \Sigma$, we can deform this $C^0$-class extension to be smooth.
Then  $\Sigma \cup \bar\Sigma$, constructed in the previous section, is achieved as the limit $\epsilon \to 0$.

\subsection{The first step}
\label{first}

The existence of a smooth inward extension from $S_\alpha$ was imposed as one of the assumptions for the theorem. 
This means that the manifold $\Sigma$ with the metric~\eqref{nonm} is extended to the negative $r$ region  slightly. 
One can take a smooth normal coordinate~\eqref{nonm} from $S_\alpha$  in the extended region $-\hat \delta < r<0$, where $\hat \delta$ is a small positive constant. 
Note that we take the coordinate such that $\varphi k =1$ is satisfied on $S_\alpha$, 
while it is not necessary to be imposed on other $r$-constant surfaces.
The Ricci scalar in the slightly extended region is nonnegative by the assumption of the theorem. 
In this subsection, we deform  $\Sigma$ in the region $-\hat \delta < r<0$, and construct a manifold $\hat \Sigma$ 
where the Ricci scalar is strictly positive and which is glued to $\Sigma$ smoothly at $r=0$. 

Since the normal coordinate in Eq.~\eqref{nonm} is smooth, we can expand the geometrical variables based on those on $S_{\alpha}$ in a sufficiently small region $-\hat \delta < r<0$. 
Hence, since $k$ is positive on $S_\alpha$, $k$ can be positive in $-\hat \delta<r<0$ with $\hat \delta$ being sufficiently small. 
Since $r^aD_a k$ and $k_{ab}$ has upper bounds due to the smoothness of $\Sigma$, there exists a positive constant $\beta$ satisfying 
\begin{eqnarray}
&&2r^a D_a k +  k^2 + k_{ab}^2 < \beta k^2 \label{defbeta}
\end{eqnarray}
everywhere in $-\hat \delta<r<0$. 
Similarly, due to the positivity of the lapse function $\varphi$ as commented after Eq.~\eqref{nonm}, there exists a positive constant $\hat \beta$ satisfying
\begin{eqnarray}
&&\varphi^{-1} k> \hat \beta k^2 . \label{defhatbeta}
\end{eqnarray}
With such sufficiently small $\hat\delta$, we introduce a metric in the region $-\hat \delta < r<0$, 
\begin{eqnarray}
&&d\hat s ^2 = \hat  \varphi^2 d r^2 + g_{ab} dx^a dx^b, 
 \label{hatm} \\
 &&
\hat \varphi:= u(r) \varphi, \\
&&u(r) := 1 - \exp\left(\frac{C \hat  \delta}{r }\right),
\end{eqnarray}
where $C$ is a constant satisfying $C>\hat \delta \beta/\hat \beta$. 
We shall call this manifold $\hat \Sigma$. 
Note that $u(r)-1$ and its any order derivatives approach zero in the limit where $r$ goes to zero from negative. 
Therefore, at $r=0$, $\hat \Sigma$ is smoothly glued to $\Sigma$. 

Let us focus on the Ricci scalar of $\hat \Sigma$. 
The $(n-1)$-dimensional geometrical variable are calculated as
\begin{eqnarray}
&&{}^{(n-1)}\hat R = {}^{(n-1)} R , \qquad \hat \varphi^{-1} \hat{\cal D}^2 \hat \varphi =  \varphi^{-1} {\cal D}^2  \varphi , \nonumber \\
&&\hat k_{ab} = u^{-1} k_{ab}, \qquad \hat k = u^{-1} k, \qquad \hat{r}^a \hat{D}_a \hat{k} = - u^{-2} \varphi^{-1} k (\partial_{r} \log u) + u^{-2} r^aD_a k,
\end{eqnarray}
where variables with hat indicate those with respect to metric~\eqref{hatm}. 
Then the $n$-dimensional Ricci scalar becomes
\begin{eqnarray}
{}^{(n)}\hat R &=&  {}^{(n-1)}\hat R - 2 \hat \varphi^{-1} \hat{\cal D}^2 \hat \varphi - 2 \hat{r}^a \hat{D}_a \hat{k} -  \hat k^2 -  \hat k_{ab}^2 
\nonumber \\ 
&=& {}^{(n)}R +\left(1-u^{-2}\right)  \left(  2 r^a D_a k +  k^2 + k_{ab}^2\right) + 2 u^{-2} \varphi^{-1} k\partial_{r} \log u.
\label{nhatRicci}
\end{eqnarray}
We can estimate the bound of the functions written in $u$,  
\begin{eqnarray}
0&>& 1-u^{-2}  
\ =\ 
u^{-2} \left[-2 \exp\left( \frac{C\hat \delta}{r}\right) + \exp\left(\frac{ 2 C\hat \delta}{r}\right) \right]  \nonumber \\
&>& -2 u^{-2} \exp\left( \frac{C\hat \delta}{r}\right),
\end{eqnarray}
and
\begin{eqnarray}
\partial_{r} \log u &=& u^{-1}  \frac{C\hat \delta}{r^2} \exp\left(\frac{C\hat \delta}{r}\right)
\ >\ 
\frac{C}{\hat \delta} \exp\left( \frac{C\hat \delta}{r}\right). \label{derrlogu}
\end{eqnarray}
For the last inequality in Eq.~\eqref{derrlogu}, we used $0<u<1$ and $-\hat\delta<r<0$. 
Through these estimates with Eqs.~(\ref{defbeta}) and (\ref{defhatbeta}), Eq. \eqref{nhatRicci} implies 
\begin{eqnarray}
{}^{(n)}\hat R \ >\ {}^{(n)}R + 2 \frac{k^2}{u^2} \left(\frac{C\hat  \beta}{\hat \delta} - \beta\right) \exp\left( \frac{C\hat \delta}{r}\right) .
\end{eqnarray}
Therefore, since the constant $C$ satisfies $C>\hat \delta \beta/\hat \beta$, 
${}^{(n)}\hat R $ is strictly positive in $-\hat \delta < r <0$. 

\subsection{The second step}
\label{second}

In the previous subsection we have constructed the extended region $\hat\Sigma$, 
where the Ricci scalar ${}^{(n)}R$ is strictly positive. 
In this subsection, we construct a further extended manifold from $\hat\Sigma$.
The strict positivity of the Ricci scalar ${}^{(n)}R$ enables the gluing discussed in Sec.~\ref{SoP} 
to be deformed into the smooth one as we will explain. 
The gluing is required to be done on a surface ${\cal S}_\epsilon$ in the extended region $\hat\Sigma$, 
which is different  from $S_\alpha$. 
The manifold $\bar \Sigma$ is constructed from $S_\alpha$ in Sec.~\ref{SoP}, 
but now we need to construct a manifold from $\hat\Sigma$ based on ${\cal S}_\epsilon$.
Therefore, the obtained manifold (named $\mathring \Sigma$) is different from $\bar \Sigma$. 
In the limit where ${\cal S}_\epsilon$ approaches  $S_\alpha$, $\mathring \Sigma$ approaches  $\bar\Sigma$.
In this subsection, we will take a region in $\hat\Sigma$, construct $\mathring \Sigma$ in the same way as that 
in Sec.~\ref{SoP}, and make the gluing between $\hat\Sigma$ and $\mathring \Sigma$ smooth.

Let us introduce a positive constant $\epsilon$, such that $0< 2\epsilon < \hat \delta$ is satisfied. 
We construct a manifold $\mathring \Sigma$ in the same way as that in Sec.~\ref{SoP}, but based on 
the surface ${\cal S}_\epsilon$ at $r=-\epsilon$. 
In the region $-2\epsilon <r\le -\epsilon$ we construct a smooth gluing between $\mathring \Sigma$ and $\bar \Sigma$.

Since the metric components~\eqref{nonm} are smooth and the deformation of the metric~\eqref{hatm} is done with the manifold $\Sigma \cup \hat \Sigma$ keeping smooth, 
at $r=-\epsilon$ for enough small  $\epsilon$ the modification from the condition~\eqref{rDkineq} is 
suppressed by order $\epsilon$, that is, there exists a constant $\gamma$ satisfying
\begin{eqnarray}
{}^{(n-1)} \hat R + 2 \hat k^{-1}  \hat \D^2  \hat k -4  \hat k^{-2} \left(  \hat \D  \hat k\right)^2\ge \left[ 2 (\alpha+\gamma \epsilon) + \frac{n}{n-1}\right]  \hat k^2  
\label{rDkmod}
\end{eqnarray}
at $r=-\epsilon$ on $\hat \Sigma$. 
We take another coordinate of $\hat \Sigma$ based on $S|_{r=-\epsilon}=: {\cal S}_\epsilon$, 
\begin{eqnarray}
&& d \doublehat{s} ^2 = \doublehat{\varphi}^2 d \rho^2 + \doublehat{g}_{ab} dx^a dx^b.  
\label{dhm}
\end{eqnarray}
Here, ${\cal S}_\epsilon$ is located at $\rho=0$ where $\doublehat{g}_{ab}|_{\rho=0}= \hat{g}_{ab}|_{r=-\epsilon}$. 
$\doublehat{\varphi}$ is chosen so that $\doublehat{\varphi}=\doublehat k^{-1} (= \hat k^{-1})$ is satisfied. 
Note that the inequality~\eqref{rDkmod} still holds true for the variables with double hat, 
\begin{eqnarray}
{}^{(n-1)} \doublehat R + 2 \doublehat k^{-1}  \doublehat \D^2  \doublehat k -4  \doublehat k^{-2} \left(  \doublehat \D  \doublehat k\right)^2\ge \left[ 2 (\alpha+\gamma \epsilon) + \frac{n}{n-1}\right]  \doublehat k^2  ,
\label{rDkmod2}
\end{eqnarray}
because it is written in the $(n-1)$-dimensional geometrical variables. 

Instead of $\bar \Sigma$ given in Sec.~\ref{SoP}, we introduce another manifold $\mathring \Sigma$, whose metric is written as
\begin{eqnarray}
d{ \mathring s}^2= \mathring{\varphi}^2 d\rho^2 + \mathring{g}_{ab}dx^a dx^b, \label{ringm}
\end{eqnarray} 
where
\begin{eqnarray}
&&\mathring{\varphi}^2: = \frac{ 1-\exp \left( -\frac{n-2}{n-1} \rho_0 \right)}{1-\exp \left[ -\frac{n-2}{n-1}(\rho+\rho_0) \right]}\exp \left( \frac{2}{n-1} \rho \right) \left.  \doublehat  \varphi^2 \right|_{\rho=0},  \\
&&\mathring{g}_{ab} := \exp \left( \frac{2}{n-1}\rho \right) \left.  \doublehat{g}_{ab} (x^a) \right|_{\rho=0} 
\end{eqnarray} 
and $\rho_0$ is defined by
\begin{eqnarray}
\exp \left(\frac{n-2}{n-1}\rho_0\right) = \frac{n+ 2(n-1)\mathring{\alpha}}{2[1+ (n-1)\mathring{\alpha}]}
\label{hatr0}
\end{eqnarray}
with a constant $\mathring \alpha$.
We attempt to glue the manifold $\hat \Sigma$ to $\mathring \Sigma$ smoothly. 
At $\rho=0$, the condition~\eqref{rDkineq} on $S_\alpha$ is modified into inequality~\eqref{rDkmod2}, 
that is,  $\alpha$ is shifted to $\alpha+\gamma \epsilon$. 
For a later convenience, we consider further shift of this parameter and set a constant $\mathring \alpha$ to be
\begin{eqnarray}
\mathring{\alpha}:= \alpha + \gamma \epsilon-\mathring{\gamma} \epsilon
\end{eqnarray}
where $\mathring{\gamma}$ is a positive constant. 
This manifold $\mathring \Sigma$ depends on $\epsilon$ and
converges to $\bar \Sigma$ in the limit $\epsilon \to 0$.
Hereinafter, geometrical quantities with circles indicate that they are associated
with the metric of Eq.~\eq{ringm} and 
quantities with subscripts $\epsilon$ show those evaluated on ${\cal S}_\epsilon$ $(\rho=0)$. 

As discussed in Sec.~\ref{SoP}, geometrical variables for the metric of Eq.~\eq{ringm} satisfy
\begin{eqnarray}
&& \mathring k_{ab} = \frac1{n-1} \mathring k \mathring  g_{ab}. \qquad
\mathring k|_{{\cal S}_\epsilon} = \doublehat{\varphi}_\epsilon^{-1}= \doublehat k|_{{\cal S}_\epsilon}  \qquad
\mathring {r}^a \mathring {D}_a \mathring {k}|_{{\cal S}_\epsilon} = \mathring \alpha \mathring  k^2|_{{\cal S}_\epsilon} \nonumber \\
&&{}^{(n)}\mathring R =  {}^{(n-1)}\mathring R - 2 \mathring \varphi^{-1} \mathring {\cal D}^2 \mathring \varphi - 2 \mathring{r}^a \mathring{D}_a \mathring{k} - \frac{n}{n-1} \mathring k^2,  \nonumber \\
&& {}^{(n-1)}\mathring R = \exp \left(-\frac{2}{n-1}\rho\right) {}^{(n-1)} \doublehat  R_\epsilon,  \qquad   
 \mathring \varphi^{-1} \mathring {\cal D}^2 \mathring \varphi = \exp \left(-\frac{2}{n-1}\rho \right )  \doublehat{\varphi}_\epsilon^{-1}  \doublehat{\cal D}^2   \doublehat{\varphi_\epsilon}, 
\nonumber \\
&&
   2 \mathring{r}^a \mathring{D}_a \mathring{k} + \frac{n}{n-1} \doublehat  k^2 = \exp \left(-\frac{2}{n-1}\rho\right) \left(2 \mathring \alpha +\frac{n}{n-1} \right) \doublehat{\varphi}_\epsilon^{-2}
   \label{mrgeo}
\end{eqnarray} 
and then, we have
\begin{eqnarray}
{}^{(n)} \mathring R = \exp \left(-\frac{2}{n-1}\rho \right)\left\{ {}^{(n-1)} \doublehat  R_\epsilon - 2 \doublehat \varphi_\epsilon^{-1}  \doublehat{\cal D}^2  \doublehat\varphi_\epsilon -  \left( 2 \mathring \alpha +\frac{n}{n-1}  \right)\doublehat\varphi_\epsilon^{-2}\right\} \ge 0 . 
\end{eqnarray} 

Since on ${\cal S}_\epsilon$ some of geometrical quantities for the metric of Eq.~\eq{nonm} and Eq.~\eq{ringm} coincide,
\begin{eqnarray}
\doublehat \varphi|_{{\cal S}_\epsilon}=\mathring \varphi|_{{\cal S}_\epsilon}, \qquad \doublehat g_{ab}|_{{\cal S}_\epsilon} = \mathring  g_{ab}|_{{\cal S}_\epsilon}, \qquad \doublehat  k|_{{\cal S}_\epsilon} =\mathring k|_{{\cal S}_\epsilon},
\end{eqnarray} 
the difference between the metric components of Eq.~\eq{nonm}
and those of Eq.~\eq{ringm} can be expressed as
\begin{eqnarray}
\doublehat g_{ab} - \mathring g_{ab} = \rho^2 T \mathring g_{ab} + \rho h_{ab}, \qquad \doublehat \varphi - \mathring \varphi = 2\rho \mathring \varphi \Phi
\end{eqnarray} 
with smooth functions $T$, $h_{ab}$ and $\Phi$. 
Here, $\rho h_{ab}$ shows the traceless part
of $\doublehat g_{ab} - \mathring g_{ab} $,
and therefore, $\mathring g^{ab} h_{ab}=0$.

Let $\check\delta$ be a positive constant such that the whole region of $-\check\delta \le \rho \le0$ on $\hat\Sigma$ is included in $-2 \epsilon < r < \epsilon$. 
Then, for arbitrary $\check \epsilon$ satisfying $0<\check\epsilon<\check\delta$, 
the whole region of $-\check\epsilon \le \rho \le0$ is also included in $-2 \epsilon < r < -\epsilon$.
We introduce a manifold $\check \Sigma$ in $-\check\epsilon \le \rho \le0$ with the following metric 
\begin{eqnarray}
{d \check s}^2= \check{\varphi}^2 dr^2 + \check{g}_{ab}dx^a dx^b, \label{checkm}
\end{eqnarray} 
where
\begin{eqnarray}
&&\check \varphi = \mathring \varphi \left[ 1 + \left(\frac{d}{d\rho} F_1^2\right) \Phi \right], \\
&&\check g_{ab} = \mathring g_{ab} \left( 1 +T F_1^2\right) + F_2 h_{ab}.
\end{eqnarray} 
$F_1$ and $F_2$ are functions of $\rho$. 
Hereinafter, the geometric functions with check marks are those with respect to the metric~\eqref{checkm}. 
For $F_1=F_2=\rho$, the metric~\eqref{checkm} is reduced to Eq.~\eqref{hatm}, while for $F_1=F_2=0$ it becomes Eq.~\eqref{ringm}. 
Hence, if both $F_1$ and $F_2$ are smoothly glued to $F:=\rho$ $(\rho>0)$ at $\rho=0$ and $\tilde F:= 0$ $(\rho<-\check\epsilon)$ at $\rho=-\check\epsilon$, 
$\check \Sigma$ is smoothly glued to $\hat\Sigma$ at $\rho=0$ and $\mathring \Sigma$ at $\rho=-\check\epsilon$.
Suppose also that $\check \Sigma$ has nonnegative Ricci scalar.
In the limit $\epsilon \to 0$, the sequence of  these smooth manifolds (with nonnegative Ricci scalar) converges to $\Sigma \cup \bar \Sigma$.
Then, the smoothing is achieved. 
Therefore, what we should do is the proof of the existence of $F_1$ and $F_2$ smoothly glued to $F:=\rho$ at $\rho=0$ and $\tilde F:=0$ at $\rho=-\check\epsilon$ with $\check\Sigma$ 
satisfying the nonnegativity of  Ricci scalar.

Let us examine the Ricci scalar of metric~\eqref{checkm}.
The metric components and their derivatives are expanded as
\begin{eqnarray}
&&\check \varphi = \mathring{\varphi}|_{\rho=0} + \cO(\rho), \qquad
\check g_{ab} = \mathring{g}_{ab}|_{\rho=0} + \cO(\rho),  
\nonumber \\
&&
\check g^{ab} = \mathring{g}^{ab}|_{\rho=0} - F_2 h^{ab} + \cO\left(\rho^2 \right), 
\nonumber \\
&&
\partial_ \rho \log \check \varphi = (\partial_\rho \log \mathring{\varphi})|_{\rho=0} + (F_1^2)'' \Phi+ \cO(\rho), 
\\
&&
\partial_ \rho \check g_{ab} = (\partial_\rho \mathring{g}_{ab})|_{\rho=0} + (F_2)' h_{ab}+ \cO(\rho), 
\nonumber \\
&&
\partial_ \rho^2 \check g_{ab} = \left. \left( \partial_\rho^2 \mathring{g}_{ab}\right)\right|_{\rho=0} 
+ \mathring{g}_{ab} \left(F_1^2 \right)'' T + (F_2)'' h_{ab} + 2 (F_2)' \partial_\rho h_{ab}+ \cO(\rho), \nonumber
\end{eqnarray}
where the prime means the derivative with respect to $\rho$, that is $(F_1)' = dF_1/d\rho$, 
and $h^{ab}:= \mathring{g}^{ac} h_{cd} \mathring{g}^{db}$.
The geometrical quantities are expressed as 
\begin{eqnarray}
&&\check k_{ab} = \mathring{k}_{ab}|_{\rho=0} + \frac{(F_2)' }{2 \mathring{\varphi}|_{\rho=0}} h_{ab} + \cO(\rho), \qquad \check k =  \mathring{k}|_{\rho=0}+ \cO(\rho), \nonumber \\
&&
\check r^a \check D_a \check k = (\mathring{r}^a \mathring{D}_a \mathring{k}) |_{\rho=0} + (F_1^2)'' \left. \left[\frac{2(n-1)T-4 \mathring{\varphi} \Phi  \mathring{k}}{4 \mathring{\varphi}^2} \right]  - (F_2^2)'' \left(\frac {h_{ab}^2}{4 \mathring{\varphi}^2 }
\right)\right|_{\rho=0} + \cO(\rho), 
\nonumber \\
&&
 {}^{(n-1)} \check R = {}^{(n-1)} \mathring{R}|_{\rho=0} + \cO(\rho), \qquad \check \varphi^{-1} \check {\cal D}^2 \check \varphi =  \left( \mathring{\varphi}^{-1} \mathring{\cal D}^2 \mathring{\varphi} \right) \Big|_{\rho=0} + \cO(\rho) , \nonumber \\
  &&{}^{(n)} \check R = {}^{(n)} \mathring{R}|_{\rho=0} +\left[2(F_2^2)'' - ({F_2}')^2\right]  \left. \left( \frac1{2 \mathring{\varphi}} h_{ab} \right)^2 \right|_{\rho=0} \nonumber \\
&&\hspace{30mm}
  - \left(F_1^2\right)''
  \left.\left[\frac{2(n-1)T-4 \mathring{\varphi} \Phi  \mathring{k} }{2\mathring \varphi^2} \right]\right|_{\rho=0}
+ \cO(\rho), 
\label{checkgeo}
\end{eqnarray}
where we used $\partial_\rho\check{g}^{ab}=-\check{g}^{ac}\check{g}^{bd}\partial_{\rho}\check{g}_{cd}$,
$\partial_\rho\mathring{g}^{ab}=-\mathring{g}^{ac}\mathring{g}^{bd}\partial_{\rho}\mathring{g}_{cd}$, 
the traceless condition of $h_{ab}$, 
{\it i.e.} $\mathring{g}^{ab} h_{ab}=0$,
and $h^{ab}\partial_{\rho}\mathring{g}_{ab}=\mathring{g}^{ab}\partial_\rho h_{ab}$
that is derived from the former three formulas.
Note also that we used 
$h_{ab}\mathring{k}^{ab}=0$, which holds from $\mathring{k}_{ab}\propto \mathring{g}_{ab}$.
One can obtain the relations between each geometric variables associated with the metrics \eq{dhm} 
and \eq{ringm} by setting $F_1=F_2=\rho$ in the above equalities as follows
\begin{eqnarray}
  &&{}^{(n)}  \doublehat R |_{\rho=0} = {}^{(n)} \mathring{R}|_{\rho=0} -  \left. \left( \frac1{2 \mathring{\varphi}} h_{ab} \right)^2 \right|_{\rho=0}
  - \left.\left[\frac{2(n-1)T-4 \mathring{\varphi} \Phi  \mathring{k} -h_{ab}^2 }{\mathring \varphi^2} \right]\right|_{\rho=0} 
  \label{nonRvsmrR}
\end{eqnarray}
and 
\begin{eqnarray}
  \doublehat r^a  \doublehat D_a  \doublehat k |_{\rho=0} = (\mathring{r}^a \mathring{D}_a \mathring{k}) |_{\rho=0} +  \left. \left[\frac{2(n-1)T-4 \mathring{\varphi} \Phi  \mathring{k}-h_{ab}^2}{2 \mathring{\varphi}^2} \right]  \right|_{\rho=0} .
  \label{Eq:equality-for-doublecheck-rDk}
\end{eqnarray}
On ${\cal S}_\epsilon$, $\doublehat r^a  \doublehat D_a  \doublehat k$ is bounded from below as
\begin{eqnarray}
\doublehat r^a  \doublehat D_a  \doublehat k |_{\rho=0} &=& 
\left. \left[- \frac12 {}^{(n)}\doublehat R - \frac12 \doublehat {\mathsf{ k}}_{ab}^2 + \frac12 {}^{(n-1)}\doublehat R 
-\doublehat \varphi^{-1} \doublehat {\cal D}^2 \doublehat \varphi  - \frac{n}{2(n-1)} \doublehat k^2 \right]\right|_{\rho=0} \nonumber \\
& \ge& - \frac12 {}^{(n)}\doublehat R|_{\rho=0} - \frac12 \left( \frac{h_{ab}^2}{2 \doublehat \varphi|_{\rho=0}} \right) ^2 + ( \alpha+ \gamma \epsilon)  \doublehat k^2|_{\rho=0},
\label{Eq:inequality-for-doublecheck-rDk}
\end{eqnarray}
where $\doublehat {\mathsf k}_{ab}$ is the traceless part of $\doublehat  k_{ab}$ and 
we use the inequality~\eqref{rDkmod2}. 
Using the second and third equations in Eq.~\eq{mrgeo}
for Eqs.~\eqref{Eq:equality-for-doublecheck-rDk} and \eqref{Eq:inequality-for-doublecheck-rDk}, we have 
\begin{eqnarray}
\left. \left[\frac{2(n-1)T-4 \mathring{\varphi} \Phi  \mathring{k}-h_{ab}^2}{ \mathring{\varphi}^2} \right]  \right|_{\rho=0} \ge  
-  {}^{(n)}\doublehat R|_{\rho=0} -  \left( \frac{h_{ab}^2}{2 \doublehat \varphi|_{\rho=0}} \right) ^2 + 2\mathring \gamma \epsilon \doublehat  k^2|_{\rho=0}\ .
\label{Tphih}
\end{eqnarray}
Now, for the moment, we assume that $F_1$ and $F_2$ satisfy 
\begin{eqnarray}
\cO\left( \check \epsilon\right) < \left( F_1^2 \right)'' \le2, \qquad 4 \left( F_2^2 \right)'' - 2\left( F_2 ' \right)^2 - 3 \left( F_1^2 \right)'' \ge \cO\left( \check \epsilon\right). 
\label{conditionF}
\label{conditionF2}
\end{eqnarray} 
Then  inequality~\eqref{Tphih} with Eq.~\eq{nonRvsmrR} and the last equation of Eq.~\eq{checkgeo} gives the lower bound of ${}^{(n)}\check R$,
\begin{eqnarray}
&& {}^{(n)}\check R \ge \frac{(F_1^2)''}{2}\,\, {}^{(n)}\doublehat R|_{\rho=0} + \left[ 2(F_2^2)'' - (F_2')^2 - \frac32 (F_1^2)'' \right] \left( \frac{h_{ab}^2}{2 \doublehat \varphi|_{\rho=0}} \right) ^2 
\nonumber \\
&&\hspace{30mm}
+\left[2-\left(F_1^2\right)''\right] \mathring \gamma \epsilon \doublehat  k^2|_{\rho=0} + \cO(\rho) .
\label{ineqsemifinal}
\end{eqnarray}
Since we are considering the region $-\check \epsilon< \rho <0$, 
inequality~\eqref{ineqsemifinal} can be estimated as
\begin{eqnarray}
&& {}^{(n)}\check R \ge \mbox{min} \left( {}^{(n)}\doublehat R|_{\rho=0} ,\, 2 \mathring \gamma \epsilon \doublehat  k^2|_{\rho=0} \right) + \cO(\check \epsilon) .
\end{eqnarray}
Note that both ${}^{(n)}\doublehat R|_{\rho=0}$ and  $2 \mathring \gamma \epsilon \doublehat  k^2$ 
are strictly positive in the region $-\check \delta \le \rho \le 0$ and do not depend on $\check\epsilon$. 
Hence, 
there exists a positive constant $\check \epsilon (< \check \delta)$ such that, for any $\rho$ in $-\check \epsilon \le \rho \le 0$, ${}^{(n)}\check R \ge 0$ holds.

In sum, if there exist functions $F_1$ and $F_2$ of $\rho$ such that 
both of them are smoothly glued to the functions $F:=\rho$ at $\rho=0$ and $\tilde F:=0$ at $\rho=-\check \epsilon$ 
and satisfy inequalities of Eq.~\eqref{conditionF2}, 
the smooth gluing of $\Sigma$ with $\bar \Sigma$ is achieved as the limit of our sequence ($\epsilon \to 0$). 
The existence of such functions is shown in Appendix~\ref{App}.

\section{No existence of minimal hypersurfaces outside ${\mathscr S}_0$}\label{OM}

In application of Bray \& Lee's theorem, the minimal hypersurface ${\mathscr S}_0$ should be the outermost one
in the smooth manifold which we have constructed.  
The proof of no existence of minimal hypersurfaces outside ${\mathscr S}_0$ is basically the same as that written 
in the previous paper~\cite{Izumi:2021hlx}. 
We briefly describe the point of the proof here. 

If the outermost minimal hypersurface exists outside ${\mathscr S}_0$, they are classified into three cases, that is,  
it encloses $S_\alpha$, it exists in $r<0$ and it has intersection with $S_\alpha$. 
The first case is prohibited due to the assumption of the theorem. 
The last case is also prohibited for the following reason. 
If the outermost minimal hypersurface intersects $S_\alpha$, 
its area of the part existing $r>0$ should be larger than $\omega_{n-1} (2Gm)^{\frac{n-1}{n-2}}$ by the assumption of the theorem. 
This results in the fact that  the area of the outermost  minimal hypersurface is larger than $\omega_{n-1} (2Gm)^{\frac{n-1}{n-2}}$, 
but it is inconsistent with Bray \& Lee's theorem. 
Therefore, the first and the last cases never occur. 

Let us investigate the second case. 
We take foliations characterized by $r=$constant in the region $-\epsilon < r <0$ on $\hat\Sigma$, $\rho=$constant in the region $-\check\epsilon < \rho <0$ on $\check\Sigma$
 and $ \rho_0<\rho <-\check\epsilon $ on $\mathring \Sigma$.
They fill all region between $S_\alpha$ and ${\mathscr S}_0$ without any overlap. 
On each foliation, $k$ is strictly positive. 
Then, as we have shown in the previous paper~\cite{Izumi:2021hlx}, no minimal hypersurfaces enclosing ${\mathscr S}_0$ exist.

Therefore, ${\mathscr S}_0$ is the outermost minimal hypersurface.

\section{Completion of the proof}
\label{comp}

Now, we can apply Bray \& Lee's theorem to the smooth manifold which we have constructed in Sec.~\ref{SE}. 
The area of ${\mathscr S}_0$ is bounded as 
\begin{eqnarray}
A[{\mathscr S}_0] \le \omega_{n-1} (2Gm)^{\frac{n-1}{n-2}}. 
\end{eqnarray}
Since we have the explicit form of metric~\eqref{checkm}, 
the area of ${\cal S}_\epsilon$ is related to that of ${\mathscr S}_0$,
\begin{eqnarray}
A\left[{\mathscr S}_0\right] = \exp (-\rho_0 ) A\left[{\cal S}_\epsilon\right] 
= \left[\frac{2(1+(n-1)\mathring \alpha)}{n+2(n-1)\mathring\alpha}\right]^{\frac{n-1}{n-2}} A\left[{\cal S}_\epsilon\right],
\end{eqnarray} 
where Eq.~\eqref{hatr0} is used in the second equality. Note that, by the construction of $\check \Sigma$, area of ${\cal S}_\epsilon$ becomes the same in $\check \Sigma$ and $\hat\Sigma$. 
Since $\hat\Sigma$ is constructed by smooth deformation of $\Sigma$ and $\Sigma$ is smooth manifold, 
the difference between areas of $S_\alpha$ and ${\cal S}_\epsilon$ is order $\epsilon$, that is 
\begin{eqnarray}
A\left[{\cal S}_\epsilon\right]= A\left[S_\alpha \right] + \cO\left(\epsilon \right).
\end{eqnarray} 
Therefore, 
\begin{eqnarray}
A\left[S_\alpha \right]
\le \omega_{n-1} \left[ \frac{n+2(n-1)\alpha}{1+(n-1)\alpha} Gm\right]^{\frac{n-1}{n-2}}
 + \cO\left(\epsilon \right)
\end{eqnarray} 
is obtained.
Taking the limit $\epsilon \to 0$, we have inequality~\eqref{ineqtheorem}.

\section{Summary}
\label{Summary}

In this paper, we have shown the inequality for refined AGPSs in asymptotically flat space with nonnegative Ricci scalar whose dimension is higher than or equal to three but less than eight.  
The definition of the refined AGPS and the statement of the main theorem are shown in Sec.~\ref{defAGPS} and in Sec.~\ref{MT}, respectively. 
The inequality is a generalization of the Riemannian Penrose inequality in higher dimensions~\cite{Fischetti:2016fbh}, 
and the higher-dimensional generalization of our previous work~\cite{Izumi:2021hlx}. 

There can exist an AGPS near spatial infinity. 
As we discussed in the previous paper~\cite{Izumi:2021hlx}, $\alpha\to \infty$ corresponds to the limit where the AGPS 
approaches  the outermost minimal hypersurface ({\it i.e.} $k=0$), 
and then our inequality is reduced to the Riemannian Penrose inequality~\cite{J&W,H&I,Bray,Fischetti:2016fbh}. 
Other limit $\alpha \to -\frac{1}{n-1}$ is achieved as the $S^{n-1}$ surface at spatial infinity or $(n-1)$-dimensional sphere in the $n$-dimensional flat space, that is, 
it corresponds to the region without gravitational field. 
Therefore, our inequality gives a relation between strong and weak gravity regions through the setting of the parameter $\alpha$. 
The properties of surfaces in strong gravity regions such as an event horizon, on the one hand, are expected to give information of quantum gravity through the black hole thermodynamics. 
On the other hand, in weak gravity region, we can use the Newtonian approximation, which may give us intuitive understanding of the area inequality. 

Furthermore, it would be interesting to explore the refined AGPS in terms of geodesics. 
In black hole observation,  a photon sphere, which is a set of the circular photon orbits in static and spherically symmetric spacetime, is an important surface~\cite{Claudel:2000yi}. 
It is defined by the behavior of null geodesics and its generalizations are introduced~\cite{Yoshino:2017gqv,Siino:2019vxh,Yoshino:2019dty,Cao:2019vlu,Yoshino:2019mqw,Siino:2021kep}, and in some of them 
the relation to the LTS, which is the version of the AGPS with $\alpha=0$, is discussed.
Recently, a nontrivial behavior of null geodesics near infinity has been reported~\cite{Amo:2021gcn,Amo:2021rxr,Amo:2022tcg}. 
Circular photon orbits can be realized in asymptotic region of spacetime in a short interval of time. 
The relation between weak and strong gravitational regions in our inequality may give a hint of the understanding of 
this temporary circular photon orbits. 

There are versions of the Riemannian Penrose inequality including the effect of the electric charge~\cite{Weinstein:2004uu,Khuri:2013ana,Khuri:2014wqa} and the angular momenta~\cite{Anglada:2017ryp,Anglada:2018czw,Dain:2017jkj,Anglada:2016dbu,Jaracz:2018jrp,Kopinski:2019von}. 
Studies of AGPSs with such contributions are partially done based on the inverse mean curvature flow~\cite{Lee:2020pre,Lee:2021hft,Lee:2022fmc}. 
Analysis with the conformal flow may give further interesting results, which is left for future works.

\section*{Acknowledgement}

K.~I. and T.~S. are supported by Grant-Aid for Scientific Research from Ministry of Education,
Science, Sports and Culture of Japan (Nos. JP17H01091, JP21H05182). K.~I., T.~S. and H.~Y. 
are also supported by JSPS(No. JP21H05189). K.~I. is also supported by JSPS Grants-in-Aid fo
13
Scientific Research (B) (JP20H01902) and JSPS Bilateral Joint Research Projects (JSPS-DST collaboration) (JPJSBP120227705). 
T.~S.
is also supported by JSPS Grants-in-Aid for Scientific Research (C) (JP21K03551). 
H.~Y. is in part supported by JSPS KAKENHI Grant Numbers
JP22H01220, and is partly supported by Osaka Central Advanced Mathematical Institute (MEXT
Joint Usage/Research Center on Mathematics and Theoretical Physics JPMXP0619217849). 

\appendix
\section{An example of function for smooth extension}
\label{App}

The proof requires the existence of functions $F_1$ and $F_2$ that satisfy Eq.~\eqref{conditionF2} in the range $-\check \epsilon <x <0$. 
(In this appendix, we use $x$ as the argument of function, instead of $\rho$.) 
We present an example. 
We decompose the range $-\check \epsilon <x <0$ into two parts,  $-\check \epsilon^2 <x <0$ and  $-\check \epsilon <x <-\check \epsilon^2$. 
In the first part , we set $F_2=x$ and smoothly glue $F_1$ to $F_1=x$ at $x=0$ and to $F_1=0$ at $x= -\check \epsilon^2$. 
Then, in this region, Eq.~\eqref{conditionF2} becomes 
\begin{eqnarray}
\cOb {\check \epsilon} < \left( F_1^2 \right)'' \le2.
\label{F1}
\end{eqnarray}
In the second part, since $F_1$ is already glued to $F_1=0$, keeping this property of $F_1$, we smoothly glue $F_2$ to $F_2=x$ at $x= -\check \epsilon^2$ and to $F_2=0$ at $x= -\check \epsilon$. 
Then, Eq.~\eqref{conditionF2} becomes 
\begin{eqnarray}
2 \left( F_2^2 \right)'' - \left( F_2 ' \right)^2  \ge \cOb{\check\ep}.
\label{F2}
\end{eqnarray}
We construct the smooth gluing for $F_1$ in $-\check \epsilon^2 <x <0$ and for $F_2$ in  $-\check \epsilon <x <-\check \epsilon^2$ in order, as shown in \ref{A.2} and in \ref{A.3} respectively. 
Before that, we first show a generic discussion to construct smooth gluing in \ref{SCP}.

\subsection{Smoothing at gluing point}
\label{SCP}
We consider a smooth gluing of functions $f_1(x)$ and $f_2(x)$ at $x=x_1$. 
Suppose that at $x=x_1$, these functions coincide with each other up to the first derivative, 
\begin{eqnarray}
f_1(x_1)=f_2(x_1), \qquad f_1'(x_1)=f_2'(x_1). 
\label{f1f2cond}
\end{eqnarray}
We define $f_3$ as $f_3:=f_1-f_2$, and then, the above conditions becomes
\begin{eqnarray}
f_3(x_1)=0, \qquad f_3'(x_1)=0. 
\end{eqnarray}
The smooth gluing of functions $f_1(x)$ and $f_2(x)$ at $x=x_1$ are equivalent to that of $f_3$ and 0. 

Let us construct a smooth function $f$ which is glued at $x=x_1$ to $0$ defined in $x<x_1$ and at $x_1+\Delta x$ to $f_3$ defined in $x>x_1+\Delta x$. 
We shall introduce a function $w$,
\begin{eqnarray}
&&w(y(x)) = \frac{1}{e^y+1}, \label{defw}\\
&&y(x) =  \frac{\Delta x}{x-x_1} + \frac{\Delta x}{x-x_1-\Delta x} .\label{defy}
\end{eqnarray}
We can show by induction for $m\ge 1 \, (m\in \mathbb{N})$
\begin{eqnarray}
\frac{d^m w}{dy^m} \in {\cal F} := \left\{ w(w-1) p (w) | p(w)\mbox{ is a polynomial of }w \right\} .
\label{calF}
\end{eqnarray}
For $m=1$, since we have 
\begin{eqnarray}
\frac{d w}{dy} = w(w-1),
\end{eqnarray}
we see that \eqref{calF} holds. 
Suppose that \eqref{calF} holds for $m=q$, that is, with a polynomial $p_q(w)$ the $q$-th order derivative of $w$ is written as
\begin{eqnarray}
\frac{d^q w}{dy^q} = w(w-1) p_q(w).
\end{eqnarray}
Differentiating it, we have 
\begin{eqnarray}
\frac{d^{q+1} w}{dy^{q+1}}= w(w-1)\frac{d }{dw}\left( w(w-1) p_q(w) \right). 
\end{eqnarray}
Since $p_q(w)$ is a polynomial, $\frac{d }{dw}\left( w(w-1) p_q(w) \right)$ is also. 
Therefore \eqref{calF} holds for $m=q+1$. 

Next, we show that $\frac{d^m w}{dx^m} $, for any $m\ge1$, asymptotes to zero in both limit $x\to x_1$ and $x\to x_1+\Delta x$. 
Let us investigate the limit $x\to x_1$ first.
From Eq.~\eqref{defw} we know 
\begin{eqnarray}
0<w<1.
\end{eqnarray}
In the neighborhood of $x=x_1$, $w$ is estimated as 
\begin{eqnarray}
|w| = \frac{1}{e^y+1} < e^{-y} = \exp\left( -\frac{\Delta x}{x-x_1} -\frac{\Delta x}{x-x_1-\Delta x} \right) 
< \bar c \exp\left( -\frac{\Delta x}{x-x_1} \right),
\end{eqnarray}
where $\bar c$ is a constant. 
Then we find that 
\begin{eqnarray}
\left|\frac{d^k w}{dy^k} \right| =\left| w(w-1) p_k(w) \right| < c_k \exp\left( - \frac{\Delta x}{x-x_1}  \right), \label{wk}
\end{eqnarray}
where $c_k$ is a constant, and we use $0<w<1$, {\it i.e.}, $w$ is bounded.
The $m$-th derivative of $w$ with respect to $x$ is expressed in terms of Fa\`{a} di Bruno's formula
\begin{eqnarray}
\frac{d^m w}{dx^m} = \sum_{k=1}^m \frac{d^k w}{dy^k} B_{m,k} \left(\frac{dy}{dx},\frac{d^2y}{dx^2}, \cdots, \frac{d^{m-k+1}y}{dx^{m-k+1}} \right),
\label{FdBformula}
\end{eqnarray}
where $B_{m,k} \left(\frac{dy}{dx},\frac{d^2y}{dx^2}, \cdots, \frac{d^{m-k+1}y}{dx^{m-k+1}} \right)$ is the 
Bell polynomial.
From the definition of $y$ (Eq.~\eqref{defy}), we have
\begin{eqnarray}
\frac{d^l y}{dx^l} = (-1)^{l} (l!) \left[ \frac{\Delta x}{(x-x_1)^{l+1}} + \frac{\Delta x}{(x-x_1-\Delta x)^{l+1}} \right]
\end{eqnarray}
and thus, $\left| \frac{d^l y}{dx^l}\right|$ is bounded by a polynomial of $(x-x_1)^{-1}$ in the neighborhood of $x=x_1$.
Since 
the right-hand side of Eq. \eqref{FdBformula} is finite sum, 
$\left|\frac{d^m w}{dx^m}\right|$ is bounded as
\begin{eqnarray}
\left|\frac{d^m w}{dx^m}\right|  <\left| p\left((x-x_1)^{-1}\right) \right| \exp\left( - \frac{\Delta x}{x-x_1}  \right) 
\end{eqnarray}
where $p\left((x-x_1)^{-1}\right)$ is a polynomial of $(x-x_1)^{-1}$. 
Since the right-hand side goes to zero in the limit $x\to x_1$,  we have
\begin{eqnarray}
\lim_{x\to x_1} \frac{d^m w}{dx^m} =0 . \label{limw}
\end{eqnarray}
In a similar way, we can also show that, in the limit $x\to x_1+\Delta x$, $\frac{d^m w}{dx^m}$ asymptotes to zero.

Here, we introduce a function $f:= f_3 w$ in the range $x_1<x<x_1+\Delta x$. 
Taking the $m$-th order derivative, we have 
\begin{eqnarray}
\frac{d^m}{d x^m}  f 
=  \left( \frac{d^m }{d x^m} f_3 \right) w + \sum_{k=1}^m \binom{m}{k}  \left( \frac{d^{m-k} }{d x^{m-k}} f_3 \right)\left( \frac{d^k }{d x^k} w \right).
\end{eqnarray}
Since  we find  from \eqref{limw} that, in the right-hand side of the above equation, the terms except the first one go to zero in the limit $x \to x_1$, 
we obtain 
\begin{eqnarray}
\lim_{x\to x_1} \frac{d^m}{d x^m}  f 
= 0,
\end{eqnarray}
where we use $\displaystyle\lim_{x\to x_1} w =0$.
Hence, $f$ is smoothly glued to $0$ at $x=x_1$. 
In a similar way, we can show the smooth gluing of $f$ and $f_3$ at $x=x_1+\Delta x$. 

Now, we estimate $f$ and its derivatives. 
The first derivative of $w$ becomes 
\begin{eqnarray}
\frac{dw}{dx} = - \frac{1}{\Delta x} \left[\left(\frac{\Delta x} {x- x_1}\right)^2+ \left(\frac{\Delta x}{x-x_1 - \Delta x}\right)^2 \right] w (w-1).
\end{eqnarray}
At first, we consider the range $x_1<x<x_1+ \frac{\Delta x}{3}$. 
Using the estimate of $w$
\begin{eqnarray}
(0<)w < e^{-y} &=& \exp\left( - \frac{\Delta x}{x-x_1 } - \frac{\Delta x}{x-x_1 - \Delta x}\right)\nonumber\\
&=&\exp\left[ - \frac{\Delta x}{2(x-x_1) }\right] \exp\left[ - \frac{\Delta x}{2(x-x_1) } - \frac{\Delta x}{x-x_1 - \Delta x}\right],
\end{eqnarray}
we have 
\begin{eqnarray}
\left|\frac{dw}{dx} \right| < \left| \frac{1}{\Delta x} \left[\left(\frac{\Delta x} {x- x_1}\right)^2+\frac{9}{4} \right] 
\exp\left[ - \frac{\Delta x}{2(x-x_1) }\right]   \right|<\frac{5}{\Delta x},
\end{eqnarray}
where we use $0<w<1$ and the fact that, in the range $x_1<x<x_1+ \frac{\Delta x}{3}$,
\begin{eqnarray}
 \exp\left[ - \frac{\Delta x}{2(x-x_1) } - \frac{\Delta x}{x-x_1 - \Delta x}\right] <1
\end{eqnarray}
holds. In the range $x_1+ \frac{2\Delta x}{3}<x<x_1+ \Delta x$, we can also show $\frac{dw}{dx} = \cOb{\Delta x^{-1}}$ in a similar way. 
For the case with $x_1+ \frac{\Delta x}{3}\le x\le x_1+ \frac{2\Delta x}{3}$, 
using $0<w<1$ and 
\begin{eqnarray}
(0<)\left(\frac{\Delta x} {x- x_1}\right)^2+ \left(\frac{\Delta x}{x-x_1 - \Delta x}\right)^2 \le9+9=18,
\end{eqnarray}
we can show $\frac{dw}{dx} = \cOb{\Delta x^{-1}}$. 
Therefore, in all range $x_1<x<x_1+\Delta x$, $\frac{dw}{dx} = \cOb{\Delta x^{-1}}$ holds. 
Similarly, we can also show $\frac{d^2w}{dx^2} = \cOb{\Delta x^{-2}}$.

Taking the range $x_1<x<x_1+ \Delta x$ to be short enough, that is, $\Delta x$ to be sufficiently small, 
in the neighborhood of $x=x_1$, $f_3$ is bounded as 
\begin{eqnarray}
f_3 = \frac12 \left. \left( \frac{d^2}{dx^2}f_3\right)\right|_{x=x_1} (x-x_1)^2 + \cOb{(x-x_1)^3}.
\end{eqnarray}
Suppose that the absolute value of $\left.\left( \frac{d^2}{dx^2}f_3\right)\right|_{x=x_1}$ takes a small one $\varepsilon$ but 
$\Delta x$ is set to be smaller than it. 
Then, in the range $x_1<x<x_1+ \Delta x$, we can estimate $f$ and its derivatives as
\begin{eqnarray}
&&|f|=| f_3 w| < \varepsilon \cOb{\Delta x^2} \cOb{\Delta x^0} =  \varepsilon \cOb{\Delta x^2} ,\\
&&\left|\frac{d}{dx}f \right| =\left| \left( \frac{d}{dx}f_3\right)w+ f_3 \left(\frac{d}{dx} w\right)\right| 
< \varepsilon \cOb{\delta x ^1}\cOb{\Delta x^{0}} +  \varepsilon\cOb{\delta x ^2} \cOb{\Delta x^{-1}} = \varepsilon\cOb{\Delta x} , \nonumber \\ \\
&&\left|\frac{d^2}{dx^2}f\right| =  \left|\left( \frac{d^2}{dx^2}f_3\right)w+ 2\left(\frac{d}{dx} f_3\right)\left(\frac{d}{dx} w\right) + f_3 \left(\frac{d^2}{dx^2} w\right)\right|  \nonumber  \\
&&\hspace{10mm}
<\varepsilon \cOb{\Delta x^{0}}+\varepsilon\cOb{\delta x ^1} \cOb{\Delta x^{-1}}
+\varepsilon\cOb{\delta x ^2} \cOb{\Delta x^{-2}} = \varepsilon \cOb{\Delta x^{0}}.
\label{f3w2}
\end{eqnarray}
Therefore, the smooth gluing is done, keeping $f$, $\frac{d}{dx}f$ and $\frac{d^2}{dx^2}f$ small.

\subsection{Smoothing $F_1$ in $-\check \ep^2 < x< 0$}
\label{A.2}

In this subsection, we construct a smooth gluing of $F_1$ from $x$ at $x=0$ to $0$ at $x=-\check \ep^2$.
As discussed in  the beginning of this appendix, $F_1$ should satisfy Eq.~\eqref{F1}. 
Moreover, if a $C^2$ function $\tilde F_1$ satisfies Eq.~\eqref{F1}, 
in the method that we have shown in \ref{SCP}, a $C^2$ function $F_1$ can be constructed with 
the difference up to the second order derivatives being as small as possible. 
However, in Eq.~\eqref{F1} the upper bound is exactly $2$, and thus 
at the point with $\tilde F_1''=2$, we cannot use the method of Appendix~\ref{SCP} 
because the deviation of function by the smoothing shown in Appendix~\ref{SCP} is tiny but nonzero. 

Now, we introduce $\tilde F_1$ as
\begin{eqnarray}
\tilde F_1 =
\begin{cases}
x& (-\gamma_1<x<0)\\
-\sqrt{F(x)}& (x_0 <x <-\gamma_1) \\
-\sqrt{g_1(x)} & (-A_1<x<x_0) \\
-\sqrt{B_1}& (-2\ep_1< x<-A_1) \\
-\sqrt{B_1} w|_{x_1=-3\ep_1, \, \Delta x=\ep_1}& (-3\ep_1<x<-2\ep_1) \\
0& (-4\ep_1<x < -3\ep_1)
\end{cases} 
\label{tF1}
\end{eqnarray}
where
\begin{eqnarray}
&&F(x) :=  x^2 - \alpha_1  \exp\left( X(x) \right),  \qquad 
g_1(x) = -\epsilon_1 (x+A_1)^2+B_1, \nonumber  \\
&&X(x) := \frac{\beta_1}{x+\gamma_1}, \qquad 
x_0 := \frac{\beta_1}{X_0}-\gamma_1, \qquad
X_0: =- (3+\sqrt{3}) , \nonumber \\
&&\alpha_1: =\frac{ 2(1+\ep_1)\beta_1^2}{X_0^3 (2+X_0) \exp (X_0)} , \qquad 
\beta_1: =\ep_1^3 , \qquad 
\gamma_1: = \ep_1^2, \qquad
 \ep_1= \frac14 {\check \ep^4},\\
&& A_1 := \frac{1+\ep_1}{\ep_1} \left( \gamma_1 - \frac{3+X_0}{X_0 (2+X_0)} \beta_1 \right) =: \frac{1+\ep_1}{\ep_1} \hat \gamma_1
, \nonumber \\
&&B_1:= \frac{1+\ep_1}{\ep_1} \hat \gamma_1 ^2 - (1+\ep_1) \frac{4+X_0}{X_0^3(2+X_0)^2} \beta_1^2. \nonumber 
\end{eqnarray}
Then, one can confirm that $\tilde F_1$ is $C^2$-class and satisfies Eq.~\eqref{F1} everywhere in the range $(-\check \ep^4=) -4 \ep_1<x <0$. 
At the gluing points of smooth segments \eqref{tF1} except $x= -\gamma_1$, the second derivative of $F_1$ is not equal to $2$, but  at $x= -\gamma_1$ it is.
Therefore, we cannot use the method shown in the Appendix~\ref{SCP} only at $x= -\gamma_1$. 
However, $\tilde F_1$ is smooth at this point. 
Let us show it. 
Since $\tilde F_1$ is strictly negative at $x= -\gamma_1$, if $F_1^2$ is smooth, $F_1$ is also. 
We investigate 
\begin{eqnarray}
\tilde F_1^2-x^2 =
\begin{cases}
0& (-\gamma_1<x<0)\\
- \alpha_1  \exp\left( X(x) \right)=- \alpha_1\exp\left( \frac{\beta_1}{x+\gamma_1} \right) & (x_0 <x <-\gamma_1) \ .
\end{cases} 
\end{eqnarray}
This function is well known to be smooth at $x=-\gamma_1$. 

As a result, we can construct a smooth function $F_1$ constructed by smoothing $\tilde F_1$ satisfying Eq.~\eqref{F1}.
The smoothing is done within the range $-\check \ep^2 < x< 0$.

\subsection{Smoothing $F_2$ in $-\check \ep < x< -\check \ep^2$}
\label{A.3}

Let us consider a function $\tilde F_2$ defined as 
\begin{eqnarray}
\tilde F_2 =
\begin{cases}
x& (-\gamma_2<x<0)\\
-\left(\tilde F(x)\right)^{\frac47}& (\tilde x_0 <x <-\gamma_2) \\
-\left(g_2(x)\right)^{\frac47} & (- A_2<x<\tilde x_0) \\
- B_2^{\frac47}& \left(-\frac12 \sqrt{\ep_2}< x<- A_2 \right) \\
- B_2^{\frac47} w|_{x_1=-\sqrt{\ep_2},\, \Delta x=\frac12 \sqrt{\ep_2}}& \left(- \sqrt{\ep_2}<x<-\frac12 \sqrt{\ep_2}\right) \\
0& (x < -\sqrt{\ep_2})
\end{cases} 
\end{eqnarray}
where
\begin{eqnarray}
&&\tilde F(x) :=  (-x)^{\frac74} - \alpha_2  \exp\left( \tilde X(x) \right),  \qquad 
 g_2(x) = \left(-\tilde x_0\right)^{-\frac14}\left[-\epsilon_2 (x+A_2)^2+B_2 \right], \nonumber  \\
&&\tilde X(x) := \frac{\beta_2}{x+\gamma_2}, \qquad 
\tilde x_0 := \frac{ \beta_2}{\tilde X_0}- \gamma_2, \qquad
\tilde X_0: =- (3+\sqrt{3}) , \nonumber \\
&&\alpha_2: = \left(-\tilde x_0\right)^{-\frac14} \frac{ \frac{21}{16}+2 \ep_2}{\tilde X_0^3 (2+\tilde X_0) \exp (\tilde X_0)}  \beta_2^2, \quad 
\beta_2: =\ep_2^3 , \quad 
\gamma_2: = \ep_2^2, \quad
 \ep_2=  {\check \ep^2},\\
&& A_2 = \frac{1}{2 \ep_2} \left[ \left( \frac74 +2 \ep_2 \right) \gamma_2 
- \frac{\frac{7}{16}(11+4 \tilde X_0) + 2 \ep_2 (3+\tilde X_0)}{\tilde X_0 (2+\tilde X_0)} \beta_2 \right] , \nonumber \\
&&B_2= \ep_2(\tilde x_0 +A_2 )^2 + \tilde x_0^2- \frac{ \frac{21}{16} + 2 \ep_2}{\tilde X_0^3 (2 + \tilde X_0)}
\beta_2^2
. \nonumber 
\end{eqnarray}
Then, one can confirm that $\tilde F_2$ is  $C^2$-class and satisfies Eq.~\eqref{F2}. 
In a similar discussion of the Appendix~\ref{A.2}, smoothing of $\tilde F_2$ is done and we obtain a smooth function satisfying Eq.~\eqref{F2}, which is done in the range $-\check \ep < x< -\check \ep^2$.


\end{document}